# POWERFUL SOURCE OF BREMSSTRAHLUNG X-RAY RADIATION BASED ON A HIGH CURRENT PULSED ELECTRON ACCELERATOR

*A.B. Batrakov, I.N. Onishchenko, S.I. Fedotov, E.G. Glushko, O.L. Rak, A.O. Zinchenko*
*National Science Center «Kharkiv Institute of Physics and Technology», Kharkiv, Ukraine*
*E-mail: a.batrakov67@gmail.com*

Enhancement of the bremsstrahlung X-ray radiation (BSXR) power generated by the high-current relativistic electron beam (REB) of a pulsed direct-action accelerator "TEMP-B" is being curried out to use in particular for studying the radiation resistance of walls material of radioactive waste containers. For this the magnetic field in which REB is transported in the drift chamber is taken higher at some distance before the converter to provide REB diameter on the converter the same as in the drift chamber. Scheme of power supply and creation of the required magnetic field for the REB current transportation through the whole distance from the diode to the converter is developed. Higher BSXR dose per REB current pulse at the same energy resource is obtained.

## INTRODUCTION

The development of high-current direct-action electron pulse accelerators [1] with beam currents of tens of kA and energies of up to 1 MeV in pulses of duration up to several microseconds has made it possible to use such beams for many applications [2], including modification and hardening of surfaces of various materials [3], generation of high-power microwave radiation [4], production of high-power hard bremsstrahlung X-ray radiation (BSXR) [5], etc. An important problem in implementing these applications is the delivery of a high-current relativistic electron beam (REB) to the object being processed. In particular for applications using high-power hard BSXR, its generation requires transporting a high-current REB to a converter that transforms the energy of REB electrons into the energy of BSXR. Such transportation is performed by a pulsed magnetic field of the appropriate strength, which provides both magnetic isolation of electrons in the accelerating diode (so-called magnetically isolated diode) and delivery of the high-current REB to the converter. The required magnetic field is experimentally realized [6-8]. However, due to the forced layout of the experiment [5] for measuring hard BSXR generated during the braking of the REB in the converter, the final part of the REB transportation way was carried out in a decreasing magnetic field in front of the converter. This led to angular divergence of the REB and only the axial part of the REB entered the converter. It reduced the possible power of the BSXR.

This paper presents the results of obtaining the BSXR of increased power due to increasing magnetic field in the region in front of the converter, that has been realized early [7], ensuring the delivery of the entire REB to the converter.

## 1. RADIATION-BEAM COMPLEX

For the generation of high-power hard bremsstrahlung X-ray radiation (BSXR) the high-current REB, produced by pulsed direct-action accelerator "TEMP-B" (E ≈0,5-1,0 MeV, I ≈10-25 kA, τ ≈1,5 μs) has been used by its injection on the converter [5]. To obtain a tubular REP in the magnetically isolated diode the tubular edge stainless steel cathode with diameter of 96 mm and edge width of 15 mm and a thickness of 1 mm was used, the cathode-anode gap is 25 mm.

The scheme of the radiation-beam complex is presented in Fig. 1.

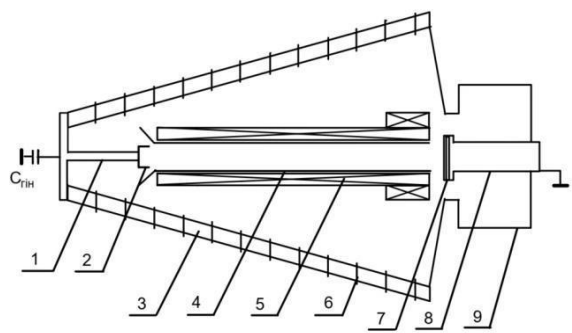

*Fig. 1. Scheme of the radiation-beam complex, consisting of magnetically isolated vacuum diode with correcting magnetic section convertor:*
*1 – cathode node; 2 – cathode; 3 – accelerator column insulator; 4 – anode insert with REB transportation chamber; 5 – main and correcting magnetic field solenoids; 6 – gradient rings of the accelerator column; 7 – convertor; 8 – X-ray input; 9 – camera for the experiments with X-ray radiation*

A method has been developed to enhance BSXR by REB current increase on the converter by means of higher magnetic field strength between the REB transportation chamber and the converter. In such a way REB current as a whole is delivered to the converter without losses.

The parameters of the magnetic system with the main and correcting solenoids have been calculated and experimentally realized [7]. The windings of the magnetic system solenoids use enameled copper wire with a diameter of 1.2 mm. The length of the REB transportation chamber on which the solenoids are located is $L_o$ =750 mm, and its outer diameter is 76 mm. The total number of turns of the main and correcting solenoid sections is N =2640. To reduce the current in the solenoid winding wire and reduce the total solenoid resistance, the main sections are connected in parallel. The correction sections are connected in series with the main sections, respectively. The arrangement of the solenoids is shown in Fig. 2.

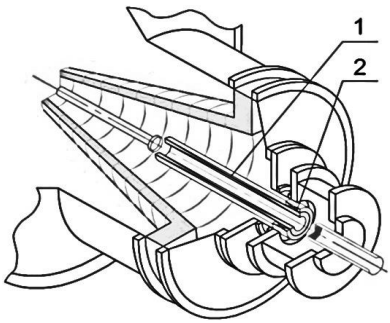

*Fig. 2. Scheme of the main (1) and correcting (2) sections of the magnetic system solenoid*

The magnetic field strength in the transportation chamber between the REB and the converter is equal to $H = 5.5 \times 10^5$ A/m for the solenoid current I = 150 A.

A capacitor bank with a capacitance of 4000 μF, charged to a voltage of 750 V, is used as a power source for the solenoid. The energy of such a charged battery of 1.125 kJ is sufficient for the solenoid pulsed operation.

When discharging a 4000 μF capacitor bank into a solenoid with an inductance of 12.7 mH and a resistance of 3.3 Ω, so an aperiodic discharge takes place as the inequality holds $r \geq 2\sqrt{L/C}$, where $r$ is the active resistance of the circuit, $L$ is the inductance of the solenoid, and $C$ is the capacitance of the capacitor bank.

Fig. 3 shows experimentally measured data on the longitudinal magnetic field intensity distribution along the axis from the cathode to the converter.

The magnetic field strength in the drift chamber and at the converter was $H = 5.5 \cdot 10^5$ A/m. It was achieved by using corrective sections, which provided at maximum magnetic field strength $H = 8.6 \cdot 10^5$ A/m (see Fig. 3).

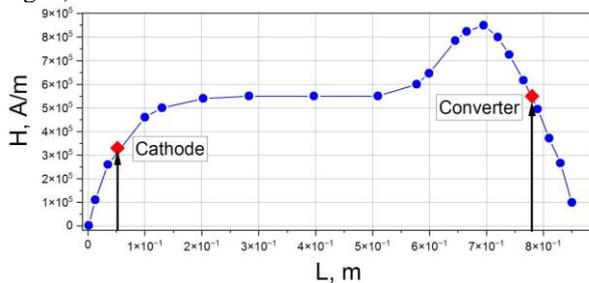

*Fig.3. Experimentally obtained distribution of the magnetic field strength along REB transportation direction*

It allows to place the convertor outside of solenoid that is convenient for the performance radiation procedure, and at the same time to where the magnetic field is the same as in the main part of the REB transportation. Thus magnetized REB, maintaining its transverse dimensions, entirely without losses reaches the converter. The REB current that was delivered to the converter was 12.5 kA, with an energy of 750 keV.

## 2. EXPERIMENTAL RESULTS OF BSXR POWER ENHANCEMENT

The power and hardness of BSXR are determined by the energy and current of the REB, the design and material of the converter. When using modern high-current pulsed accelerators, only 4-6 % of the electron energy is converted into BSXR. As a result of the thermal effect of the high-current REB on the converter material, it is deformed, cracked, partially or completely melted and destroyed, that requires its replacement and extends the time of experimental research. Therefore, the choice of the converter material is reduced to its stability under the action of REB at a minimum thickness and maximum efficiency of converting REB energy into BSXR.

When choosing a converter design, its mechanical and thermophysical properties are important, which must ensure high stability and durability under the influence of extremely high pulsed mechanical and thermal loads under the action of high-current REB.

Fig. 4 shows the REB imprint on a stainless steel converter obtained on a radiation-beam complex using a tubular cathode.

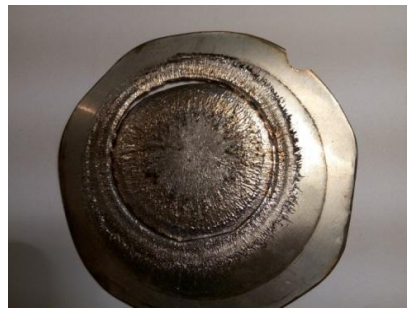

*Fig. 4. Tubular REP imprint on a stainless steel converter after 5 pulses*

From the imprint with a diameter of 55 mm and a ring width of 5 mm, shown in Fig. 4, it is possible to make a conclusion about the transverse size and wall thickness of the tubular REB before the convertor.

From Fig. 4 it is seen that the converter made of steel is quickly being destroyed under the action of the injected REB with a current of 21 kA and an energy of 0.75 MeV. After 5 pulses, 2 mm of the converter was burned. Therefore, steel should not be used as a material for the converter.

When choosing a material for the target, not only the BSXR intensity is taken into account, which increases quadratically with an increase in the charge of the nuclei of the substance, but also the prevalence and cost of the material and its physical properties, the main of which is the melting point, since most of the electron energy is converted into heat when interacting with the target. For reasons of heat resistance, tantalum with Z =73 or tungsten with Z =74 are used as target materials. The melting point of tungsten is 3380°K.

For the convenience of performing experiments, vacuum X-ray channels were developed and created, which allow changing samples for irradiation without disturbing the vacuum conditions, as well as simultaneously measuring the parameters of BSXR. Two vacuum inlets with a diameter of 50 mm and a length of 390 mm and a diameter of 75 mm and a length of 500 mm were manufactured.

The scheme of the experiment and the converter is shown in Fig. 5.

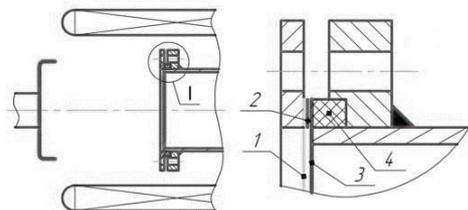

*Fig. 5. Scheme of the experiment and the converter*

A cylindrical edge cathode with a diameter of 96 mm, an edge width of 15 mm and an edge thickness of 1 mm was used. The cathode-anode gap is 30 mm.

A converter was designed and manufactured to obtain maximum doses of BSXR. All current-carrying elements of the converter are electrically interconnected and under the same potential. The combined converter consists of:

1. Titanium plate, 30-50 microns thick, performing the following functions: a protective shield that absorbs the shock wave, a beam filter (low-energy electrons and cathode plasma ions are deposited on the plate and diverted to the ground), a collector and a shutter for emissions from the collector directed in the opposite direction.

2. Copper washer, 200 microns thick, which provides the necessary distance between the screen and the converter, playing the role of a damper, and is also used to relieve thermal load.

3. A converter made of tantalum with a thickness of 50 to 500 microns, in which the main process of converting electron beam energy into thermal energy and braking radiation takes place.

4. Gaskets made of vacuum rubber, 4 mm thick, which acts as an shock wave absorber and vacuum insulator.

In general, this design reduces the energy load on the converter, which increases its service life by 30 %.

Sampless were irradiated as follows: an electron beam formed in the magnetically isolated vacuum diode with was transported through the drift chamber to the converter for generation of BSXR, which was recorded behind the converter.

Dose measurements of BSXR were performed using LiF-based thermo-luminescent sensors. placed directly behind the converter at different distances to determine the spatial distribution of the BSXR. Each dose value was obtained by averaging a series of measurements from five pulses. The overall dimensions of the detector are $5 \times 5 \times 15$ mm, its resistance is $\sim 10^9\ \Omega$. The signal from the output of the semiconductor detector is the current resulting from ionization in the detector, which has a linear characteristic with respect to the dose rate and does not depend on temperature.

The placement scheme of thermoluminescent sensors for BSXR measurement is shown in Fig. 6

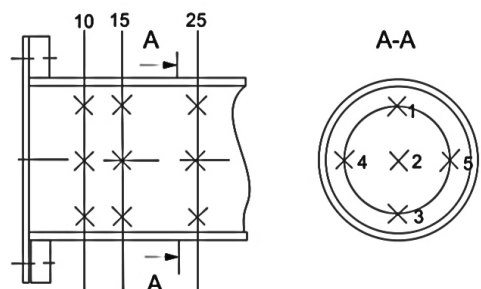

*Fig. 6. Placement scheme of sensors*

Results of measuring absorbed BSXR doses obtained with a tubular REB parameters: - current 21.5 kA, energy 750 keV, pulse duration 1.5 μs, from sensors located at different distances from the converter are presented in Table 1.

*Table 1.*

| Number of sensors | Distance from the converter, m | | |
|---|---|---|---|
| | $1 \cdot 10^{-2}$ | $1,5 \cdot 10^{-2}$ | $2,5 \cdot 10^{-2}$ |
| #1, dose, R | $9,1 \cdot 10^3$ | $9,3 \cdot 10^3$ | $9,8 \cdot 10^3$ |
| #2, dose, R | $8,7 \cdot 10^3$ | $11,2 \cdot 10^3$ | $11,3 \cdot 10^3$ |
| #3, dose, R | $8,9 \cdot 10^3$ | $8,9 \cdot 10^3$ | $9,1 \cdot 10^3$ |
| #4, dose, R | $9,4 10^3$ | $9,4 \cdot 10^3$ | $9,9 \cdot 10^3$ |
| #5, dose, R | $8,6 \cdot 10^3$ | $9,6 \cdot 10^3$ | $9,5 \cdot 10^{3\cdot}$ |

As it can be seen from Table 1, the absorbed BSXR radiation field per pulse obtained with an extended magnetic field up to the converter is almost an order of magnitude greater than that obtained in [5] without additional correcting solenoids providing the entire REB delivery up to the converter. The maximum absorbed BSXR dose from sensor #2, which is located on the axis at a distance of 25 mm from the converter, was 11300 R for the impulse.

## CONCLUSIONS

At the radiation-beam complex using the high-current REB of direct action pulse electron accelerator "TEMP-B" the maximum possible at this complex dose per pulse and its distribution in space have been achieved. This problem has been solved by extending the REB transport path with a magnetic field of uniform intensity up to the converter. The problem of obtaining significantly higher doses can be solved by a large number of pulses, i.e. by the sequence mode of BSXR generation. The pulse repetition rate is determined by the requirements of the time of vacuum conditions recovery after each pulse, the establishment of a stationary permissible thermal load of the solenoid creating the transporting magnetic field, and the resistance to converter destruction.

## REFERENCES

1. J.W. Poukey, J.R. Freeman, G. Jonas, Simulation of Relativistic Electron Beam Diodes // *Physics of plasma*, 1995, Vol. 10. No.6, p. 954–958.
2. Yu.F. Lonin, I.I. Magda. High-current relativistic accelerators of IPENMA NSC KIPT and their applications. // *Problems of Atomic Science and*

*Technology, Series: Nuclear Physics Investigations* (50).2008, No.5, p. 85-90.
3. F. Klepikov, Yu.F. Lonin, V.V. Lytvynenko, A.V. Pashenko, A.G. Ponomarev, V.V. Uvarov, V.T. Uvarov, V.I. Sheremet. The formation of strengthening coats by microsecond // *Problems of Atomic Science and Technology, Series: Nuclear Physics Investigations* (50).2008, No.5, p. 91-95.
4. P.T. Chupikov, R.J. Faehl, I.N. Onishchenko, Yu.V. Prokopenko, S.S. Pushkarev. Vircator Efficiency Enhancement at Plasma Assistance. // *IEEE Transactionson Plasma Science*, 2006, V. 34, No. 1, p.14-17.
5. A.B. Batrakov, E.G. Glushko, A.M. Yegorov, A.A. Zinchenko, V.V. Litvinenko, Yu.F. Lonin, A.G. Ponomaryov, A.V. Rybka, S.I. Fedotov, V.T. Uvarov. Study of hard braking X-ray radiation on the radiation-beam complex «TEMP» // *Problems of Atomic Science and Technology Series. Nuclear Physics Investigations*. 2015. №6. (100), p.100-104.
6. Heinz Knoepfel, Pulsed high magnetic fields, North Holland Publising company Amsterdam. London. 1970.
7. A.B. Batrakov, V.M. Zalkind, Yu.F. Lonin, A.G. Ponomarev, V.T. Uvarov, P.T. Chupikov. Magnetic system for transportation high current electron relastivitic beam // *Problems of Atomic Science and Technology. Series: Plasma electronics and new acceleration methods*. 2008, № 4, p. 303 – 305.
8. A.B. Batrakov, E.G. Glushko, A.A. Zinchenko, Yu.F. Lonin, A.G. Ponomarev, S.I. Fedotov. Pulsed magnetic system of the relativistic electron beam accelerator «Temp-B» // *Problems of Atomic Science and Technology. Series: Plasma electronics and new acceleration methods* (8). 2013, № 4 (86), p. 7 – 9.

# ПОТУЖНЕ ДЖЕРЕЛО ГАЛЬМІВНОГО РЕНТГЕНІВСЬКОГО ВИПРОМІНЮВАННЯ НА ОСНОВІ СИЛЬНОСТРУМНОГО ІМПУЛЬСНОГО ПРИСКОРЮВАЧА ЕЛЕКТРОНІВ

*А.Б. Батраков, І.М. Оніщенко, С.І. Федотов, Є.Г. Глушко, О.Л. Рак, А.О. Зінченко*

Підвищення потужності гальмівного рентгенівського випромінювання (ГРВ), створюваного сильнострумовим релятивістським електронним пучком (РЕП) імпульсного прискорювача прямої дії «ТЕМП-Б», розробляється для використання, зокрема, в дослідженнях радіаційної стійкості матеріалу стінок контейнерів радіоактивних відходів. Для цього магнітне поле, що транспортує РЕП в транспортній камері, береться вище на деяку відстань перед конвертером, щоб забезпечити діаметр РЕП на конвертері такий же, як і в дрейфовій камері. Розроблено схему живлення та створення необхідного магнітного поля для транспортування струму РЕП на всій відстані від діода до перетворювача. Отримується більш висока доза BSXR на імпульс струму РЕП при тому ж енергетичному ресурсі.